\RequirePackage{lineno}
\documentclass[aps,twocolumn,showpacs,byrevtex]{revtex4}
\usepackage{times}
\usepackage{amsmath}
\usepackage{graphicx}
\usepackage{color}
\newcommand{\z}{Z_c(3900)}

\newcommand{\pp}{\pi^+\pi^-}

\newcommand{\LL}{\ell^+\ell^-}
\newcommand{\EE}{e^+e^-}
\newcommand{\MM}{\mu^+\mu^-}
\newcommand{\TT}{\tau^+\tau^-}
\newcommand{\GG}{\gamma\gamma}

\newcommand{\jpsi}{J/\psi}

\newcommand{\psithr}{\psi(4040)}
\newcommand{\psifou}{\psi(4160)}
\newcommand{\psifiv}{\psi(4415)}
\newcommand{\ppjpsi}{\pi^+\pi^-J/\psi}
\newcommand{\BR}{\mathcal{B}}

\begin{document}

\title{\boldmath
Precise measurement of the $e^+e^-\to \pi^+\pi^-J/\psi$ cross section
at center-of-mass energies from 3.77 to 4.60~GeV}

\author{
  M.~Ablikim$^{1}$, M.~N.~Achasov$^{9,e}$, S. ~Ahmed$^{14}$,
  X.~C.~Ai$^{1}$, O.~Albayrak$^{5}$, M.~Albrecht$^{4}$,
  D.~J.~Ambrose$^{44}$, A.~Amoroso$^{49A,49C}$, F.~F.~An$^{1}$,
  Q.~An$^{46,a}$
% ...
\\
\vspace{0.2cm}
(BESIII Collaboration)\\
\vspace{0.2cm} {\it
$^{1}$ Institute of High Energy Physics, Beijing 100049, People's Republic of China\\
% ...
$^{4}$ Bochum Ruhr-University, D-44780 Bochum, Germany\\
$^{5}$ Carnegie Mellon University, Pittsburgh, Pennsylvania 15213, USA\\
% ...
$^{9}$ G.I. Budker Institute of Nuclear Physics SB RAS (BINP), Novosibirsk 630090, Russia\\
% ...
$^{14}$ Helmholtz Institute Mainz, Johann-Joachim-Becher-Weg 45, D-55099 Mainz, Germany\\
% ...
$^{44}$ University of Rochester, Rochester, New York 14627, USA\\
% ...
$^{46}$ University of Science and Technology of China, Hefei 230026, People's Republic of China\\
% ...
$^{49}$ (A)University of Turin, I-10125, Turin, Italy; (B)University of Eastern Piedmont, I-15121, Alessandria, Italy; (C)INFN, I-10125, Turin, Italy\\
% ...
$^{a}$ Also at State Key Laboratory of Particle Detection and Electronics, Beijing 100049, Hefei 230026, People's Republic of China\\
% ...
$^{e}$ Also at the Novosibirsk State University, Novosibirsk, 630090, Russia\\
% ...
}
}

\date{\today}

\begin{abstract}

The cross section for the process $\EE\to \ppjpsi$ is measured precisely  at center-of-mass energies
from 3.77 to 4.60~GeV using 9~fb$^{-1}$ of
data collected with the BESIII detector operating at the BEPCII
storage ring. Two resonant structures are observed in a fit
to the cross section. The first
resonance has a mass of $(4222.0\pm 3.1\pm 1.4)$~MeV/$c^2$ and a
width of $(44.1\pm 4.3\pm 2.0)$~MeV, while the second one has a
mass of $(4320.0\pm 10.4 \pm 7.0)$~MeV/$c^2$ and a width of
$(101.4^{+25.3}_{-19.7}\pm 10.2)$~MeV, where the first errors are
statistical and second ones are systematic. The first resonance 
agrees with the $Y(4260)$
resonance reported by previous experiments. The precision of its
resonant parameters is improved significantly. 
The second resonance is observed in $\EE\to \ppjpsi$
for the first time. The statistical significance of this resonance is estimated to
be larger than $7.6\sigma$. The mass and width of the second resonance
agree with the $Y(4360)$ resonance reported by the $BABAR$ and Belle 
experiments within errors. Finally, the $Y(4008)$ resonance previously observed by the Belle
experiment is not confirmed in the description of the BESIII data.

\end{abstract}


\maketitle

The process $\EE\to \ppjpsi$ at center-of-mass (c.m.) energies
between 3.8 and 5.0~GeV was first studied by the \emph{BABAR}
experiment using an initial-state-radiation (ISR)
technique~\cite{babary}, and a new structure, the $Y(4260)$, was
reported with a mass around 4.26~GeV/$c^2$. This observation was
immediately confirmed by the CLEO~\cite{cleoy} and Belle
experiments~\cite{belley} in the same process. In addition, the
Belle experiment reported an accumulation of events at around 4~GeV,
which was called $Y(4008)$ later. Although the $Y(4008)$ state is
still controversial --- a new measurement by the \emph{BABAR}
experiment does not confirm it~\cite{babarnew}, while an updated
measurement by the Belle experiment still supports its
existence~\cite{bellenew} --- the observation of the $Y$-states
has stimulated substantial theoretical discussions on their
nature~\cite{epjc-review,zhusl_review}.

Being produced in $\EE$ annihilation, the $Y$-states have quantum
numbers $J^{PC}=1^{--}$. However, unlike the known $1^{--}$
charmonium states in the same mass range, such as $\psithr$,
$\psifou$ and $\psifiv$~\cite{pdg}, which decay predominantly into 
open charm final states [$D^{(*)}\bar{D}^{(*)}$], the $Y$ states
show strong coupling to hidden-charm final states~\cite{hidden}.
Furthermore, the observation of the states $Y(4360)$ and $Y(4660)$ in
$\EE\to\pp\psi(2S)$~\cite{y4360}, together with the
newly observed resonant structures in $\EE\to\omega\chi_{c0}$~\cite{omega-cc0} 
and $\EE\to\pp h_c$~\cite{pphc},
overpopulate the vector
charmonium spectrum predicted by potential
models~\cite{potential}. All of this indicates that the $Y$ states
may not be conventional charmonium states, and they are good
candidates for new types of exotic particles, such as hybrids,
tetraquarks, or meson molecules~\cite{epjc-review,zhusl_review}.

The $Y(4260)$ state was once considered a good hybrid
candidate~\cite{hybrid} since its mass is close to the value
predicted by the flux tube model for the lightest hybrid
charmonium~\cite{flux-tube}. Recent lattice calculations
also show a $1^{--}$ hybrid charmonium could have a mass
of $4285\pm 14$~MeV/$c^2$~\cite{hybrid-lattice} or
4.33(2)~GeV/$c^2$~\cite{hybrid-y4360}. Meanwhile, the
diquark-antidiquark tetraquark model predicts a wide spectrum of states
which can also accommodate the $Y(4260)$~\cite{tetraquark}.
Moreover, the mass of $Y(4260)$ is near the mass threshold
of $D_s^{*+}D_s^{*-}$, $\bar{D}D_1$, $D_0\bar{D}^*$ and 
$f_0(980)\jpsi$, and $Y(4260)$ was supposed to be a meson molecule 
candidate of these meson pairs~\cite{molecule,molecule-more}.
A recent observation of a charged charmoniumlike state
$\z$ by BESIII~\cite{zc3900}, Belle~\cite{bellenew}  and with CLEO data~\cite{zc-cleo}
seems favor the $\bar{D}D_1$ meson pair option~\cite{molecule}.
Another possible interpretation describes the $Y(4260)$ as a heavy charmonium ($\jpsi$) 
being bound inside light hadronic matter --- hadro-charmonium~\cite{hadro-charmonium}. 
To better identify the nature of
the $Y$ states and distinguish various models, more precise
experimental measurements, including the production cross section, the mass and
width of the $Y$ states, are essential.

In this Letter, we report a precise measurement of the $\EE\to
\ppjpsi$ cross section at $\EE$ c.m. energies from 3.77 to
4.60~GeV, using a data sample with an integrated luminosity of
$9.05$~fb$^{-1}$~\cite{lum} collected with the BESIII detector
operating at the BEPCII storage ring~\cite{bes3-detector}. The
$\jpsi$ candidate is reconstructed with its leptonic decay modes
($\MM$ and $\EE$). The data sample used in this measurement
includes two independent data sets. A high luminosity data set
(dubbed ``XYZ data") contains more than 40~pb$^{-1}$ at
each c.m.\ energy with a total integrated luminosity of 8.2~fb$^{-1}$,
which dominates the precision of this measurement; 
and a low luminosity data set (dubbed ``Scan data") contains 
about 7--9~pb$^{-1}$ at each c.m.\ energy with a total integrated luminosity
of 0.8~fb$^{-1}$. The integrated luminosities are measured with 
Bhabha events with an uncertainty of $1\%$~\cite{lum}. 
The c.m.\ energy of each data set is measured using 
dimuon events, with an uncertainty of $\pm 0.8$~MeV~\cite{ecm}.

The BESIII detector is described in detail elsewhere~\cite{bes3-detector}. 
%
The {\sc geant4}-based~\cite{geant4} Monte Carlo (MC) simulation
software package {\sc boost}~\cite{boost}, which includes the
geometric description of the BESIII detector and the detector
response, is used to optimize event selection criteria, determine
the detection efficiency, and estimate the backgrounds. For the
signal process, we generate 60,000 $\EE\to \ppjpsi$ events at each
c.m.\ energy of the ``XYZ data", and an extrapolation is performed to
the ``Scan data" with nearby c.m.\ energies. At $\EE$ c.m.\ energies between
4.189 and 4.358~GeV, the signal events are generated according to the
Dalitz plot distributions obtained from the data set at corresponding c.m. energy, since
there is significant $\z$ production~\cite{bellenew,zc3900,zc-cleo}. At
other c.m. energies, signal events are generated using an {\sc
evtgen}~\cite{evtgen} phase space model. The $\jpsi$ decays into
$\MM$ and $\EE$ with same branching fractions~\cite{pdg}. The ISR is
simulated with {\sc kkmc}~\cite{kkmc}, and the maximum ISR photon
energy is set to correspond to a $3.72$~GeV/$c^2$ production
threshold of the $\ppjpsi$ system. Final-state-radiation (FSR) is
simulated with {\sc photos}~\cite{photos}. Possible background
contributions are estimated with {\sc kkmc}-generated inclusive MC
samples [$\EE\to\EE,~\MM,~\TT,~\GG,~\gamma_{ISR}\jpsi,~\gamma_{ISR}\psi(2S),$
and $q\bar{q}$ with $q=u,~d,~s,~c$] with comparable integrated luminosities to the ``XYZ data".

Events with four charged tracks with zero net charge are selected.
For each charged track, the polar angle in the drift chamber must satisfy
$|\cos\theta|<0.93$, and the point of closest approach to the
$\EE$ interaction point must be within $\pm 10$~cm in the beam
direction and within $1$~cm in the plane perpendicular to the beam
direction. Taking advantage of the fact that
pions and leptons are kinematically well
separated in the signal decay, charged tracks with momenta larger
than 1.06~GeV/$c$ in the laboratory frame are assumed to be
leptons, and the others are assumed to be pions. We use the energy
deposited in the electromagnetic calorimeter (EMC) to separate electrons from muons. For both
muon candidates, the deposited energy in the EMC is required to be
less than 0.35~GeV, while for both electrons, it is required to be
larger than 1.1~GeV. To avoid systematic errors due to unstable
operation, the muon system is not used here.
Each event is required to have one $\pp\LL$ ($\ell=e$ or $\mu$) combination.

To improve the momentum and energy resolution and to reduce the
background, a four-constraint (4C) kinematic fit is applied to the event
with the hypothesis $\EE\to \pp\LL$, which
constrains the total four-momentum of the final state
particles to that of the initial colliding beams. The
$\chi^2/{ndf}$ of the kinematic fit is required to be less than $60/4$.

To suppress radiative Bhabha and radiative dimuon ($\EE\to
\gamma\EE / \gamma\MM$) backgrounds associated with photon
conversion to an $\EE$ pair which subsequently is misidentified as a $\pp$
pair, the cosine of the opening angle of the pion-pair
($\cos\theta_{\pp}$) candidates is required to be less than 0.98
both for $\jpsi\to\MM$ and $\EE$ events.
For $\jpsi\to \EE$ events, since there are more abundant photon
sources from radiative Bhabha events, we further require the
cosine of the opening angles of both pion-electron pairs
($\cos\theta_{\pi^\pm e^\mp}$) to be less than 0.98.
These requirements remove almost all of the Bhabha and dimuon
background events, with an efficiency loss of less than 1\%
for signal events.

After imposing the above selection criteria, a clear $\jpsi$ signal is
observed in the invariant mass distribution of the lepton pairs
[$M(\LL)$]. The mass resolution of the $M(\LL)$ 
distribution is estimated to be $(3.7\pm 0.2)$~MeV/$c^2$ for $\jpsi\to\MM$, and
$(3.9\pm 0.3)$~MeV/$c^2$ for $\jpsi\to\EE$ in data for the range of c.m. energies 
investigated in this study. The $\jpsi$ mass
window is defined as $3.08<M(\LL)<3.12$~GeV/$c^2$. In order to
estimate the non-$\jpsi$ backgrounds contribution, we also define the
$\jpsi$ mass sideband as $3.00<M(\LL)<3.06$~GeV/$c^2$ and
$3.14<M(\LL)<3.20$~GeV/$c^2$, which is three times as wide as the
signal region. The dominant background comes from $\EE\to q\bar{q}$
($q=u,~d,~s$) processes, such as $\EE\to \pp\pp$. Since $q\bar{q}$
events form a smooth distribution in the $\jpsi$ signal region, their
contribution is estimated by the $\jpsi$ mass sideband.
Contributions from backgrounds related with charm quark production, 
such as $\EE\to\eta\jpsi$~\cite{etaJpsi}, $D^{(*)}\bar{D^{(*)}}$ and 
other open-charm mesons, are estimated to be negligible
according to MC simulation studies.

In order to determine the signal yields, we make use of both 
fitting and counting methods on the $M(\LL)$ distribution. 
In the ``XYZ data", each data set contains many signal events,
and an unbinned maximum likelihood fit to the $M(\LL)$
distribution is performed. We use a MC simulated signal shape convolved with
a Gaussian function (with standard deviation $1.9$~MeV, which represents the resolution
difference between the data and the MC simulation) as the signal
probability density function (PDF), and a linear term for the
background. For the ``Scan data", due to the low statistics, we directly count
the number of events in the $\jpsi$ signal region and that of the
normalized background events in the $\jpsi$ mass sideband, and
take the difference as the signal yields.

\begin{figure*}
\begin{center}
\includegraphics[height=2in]{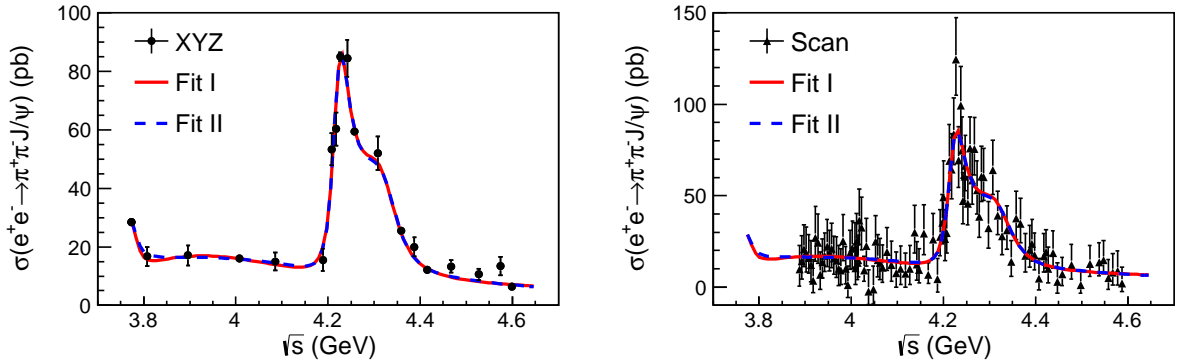}
\caption{Measured cross section $\sigma(\EE\to \ppjpsi)$ and 
simultaneous fit to the ``XYZ data" (left) and ``Scan data"
(right) with the coherent sum of three Breit-Wigner functions (red solid
curves) and the coherent sum of an exponential continuum and two Breit-Wigner
functions (blue dashed curves). Dots with error bars are data.} \label{xsec-fit}
\end{center}
\end{figure*}

The cross section of $\EE\to \ppjpsi$ at a certain $\EE$ c.m. energy $\sqrt{s}$ is calculated using
\begin{equation}
\sigma(\sqrt{s}) = \frac{N^{\rm sig}}{\mathcal{L}_{\rm int}
(1+\delta)\epsilon\mathcal{B}},
\end{equation}
where $N^{\rm sig}$ is the number of signal events,
$\mathcal{L}_{\rm int}$ is the integrated luminosity of data,
$1+\delta$ is the ISR correction factor, $\epsilon$ is the
detection efficiency, and $\mathcal{B}$ is the branching fraction
of $\jpsi\to \LL$~\cite{pdg}.
The ISR correction factor is calculated using the {\sc kkmc}~\cite{kkmc} program.
To get the correct ISR photon energy distribution, we use the $\sqrt{s}$ 
dependent cross section line shape of the $\EE\to\ppjpsi$ process, i.e. $\sigma({\sqrt{s}})$ to
replace the default one of {\sc kkmc}. Since $\sigma({\sqrt{s}})$ is what
we measure in this study, the ISR correction procedure needs to be iterated,
and the final results are obtained when the iteration converges. 
Figure~\ref{xsec-fit} shows the measured cross section $\sigma({\sqrt{s}})$ 
from both the ``XYZ data" and ``Scan data" 
(Numerical results are listed in the supplemental material \cite{supplement}).

To study the possible resonant structures in the $\EE\to \ppjpsi$
process, a binned maximum likelihood fit is performed simultaneously to the
measured cross section $\sigma({\sqrt{s}})$ of the ``XYZ data" with Gaussian uncertainties 
and the ``Scan data" with Poisson uncertainties.
The PDF is parameterized as the
coherent sum of three Breit-Wigner (BW) functions, together with
an incoherent $\psi(3770)$ component which accounts for the decay of
$\psi(3770)\to\ppjpsi$, with $\psi(3770)$ mass and width fixed to
PDG~\cite{pdg} values. Due to the lack of data near the
$\psi(3770)$ resonance, it is impossible to determine the relative
phase between the $\psi(3770)$ amplitude and the other amplitudes.
The amplitude to describe a resonance $R$ is written as
\begin{equation}
 \mathcal{A}(\sqrt{s})=\frac{M}{\sqrt{s}}
 \frac{\sqrt{12\pi\Gamma_{\EE}\Gamma_{\rm tot}\mathcal{B}_R}}
 {s-M^2+iM\Gamma_{\rm tot}}
 \sqrt{\frac{\Phi(\sqrt{s})}{\Phi(M)}}e^{i\phi},
\end{equation}
where $M$, $\Gamma_{\rm tot}$ and $\Gamma_{\EE}$ are the mass,
full width and electronic width of the resonance $R$,
respectively; $\mathcal{B}_R$ is the branching fraction of the decay $R\to
\ppjpsi$; $\Phi(\sqrt{s})$ is the phase space factor of the three-body 
decay $R\to \ppjpsi$~\cite{pdg}, and $\phi$ is the phase of the amplitude. 
The fit has four solutions with
equally good fit quality~\cite{multi-solution} and identical masses and widths of the
resonances (listed in Table~\ref{fit-mass}), while the phases and 
the product of the electronic widths with the branching fractions are different (listed in
Table~\ref{para-fit}). Figure~\ref{xsec-fit} shows
the fit results. The resonance $R_1$ has a mass and width consistent
with that of $Y(4008)$ observed by Belle~\cite{bellenew} 
within $1.0\sigma$ and $2.9\sigma$, respectively. The
resonance $R_2$ has a mass $4222.0\pm3.1$~MeV/$c^2$, which 
agrees with the average mass, $4251\pm 9$~MeV/$c^2$~\cite{pdg}, of the $Y(4260)$ 
peak~\cite{babary,cleoy,belley,babarnew,bellenew} within $3.0\sigma$.
However, its measured width is much narrower than the
average width, $120\pm 12$~MeV~\cite{pdg}, of the $Y(4260)$. 
We also observe a new resonance $R_3$. The statistical
significance of $R_3$ is estimated to be $7.9\sigma$ (including
systematic uncertainties) by comparing the change of
$\Delta(-2\ln\mathcal{L})=74.9$ with and without the $R_3$
amplitude in the fit, and taking the change of number of degree of
freedom $\Delta ndf=4$ into account. The fit quality is estimated using a
$\chi^2$-test method, with $\chi^2/ndf=93.6/110$.
Fit models taken from previous experiments~\cite{babary,cleoy,belley,babarnew,bellenew}
are also investigated and are ruled out with a confidence level equivalent to 
more than $5.4\sigma$.

\begin{table}[htbp]
\caption{The measured masses and widths of  the resonances from
the fit to the $\EE\to\ppjpsi$ cross section with three coherent
Breit-Wigner functions. The numbers in the brackets correspond to
a fit by replacing $R_1$ with an exponential describing the continuum. The errors
are statistical only.} \label{fit-mass}
\begin{tabular}{cc}
\hline\hline
Parameters & Fit result \\
\hline
$M(R_1)$ & $3812.6^{+61.9}_{-96.6}$~($\cdots$) \\
$\Gamma_{\rm tot}(R_1)$ & $476.9^{+78.4}_{-64.8}$~($\cdots$) \\
$M(R_2)$ & $4222.0\pm 3.1$~($4220.9\pm2.9$) \\
$\Gamma_{\rm tot}(R_2)$ & $44.1\pm 4.3$~($44.1\pm 3.8$) \\
$M(R_3)$ & $4320.0\pm 10.4$~($4326.8\pm 10.0$) \\
$\Gamma_{\rm tot}(R_3)$ & $101.4^{+25.3}_{-19.7}$~($98.2^{+25.4}_{-19.6}$) \\
\hline\hline
\end{tabular}
\end{table}

\begin{table*}[htbp]
\caption{The values of $\Gamma_{\EE}\BR(R\to\ppjpsi)$ (in eV) from a
fit to the $\EE\to \ppjpsi$ cross section. $\phi_1$ and $\phi_2$
(in degrees) are the phase of the resonance $R_2$ and
$R_3$, the phase of resonance $R_1$ (or continuum) is set to 0. 
The numbers in the brackets correspond to the fit by
replacing resonance $R_1$ with an exponential to describe the continuum. The
errors are statistical only.} \label{para-fit}
\begin{tabular}{ccccc}
\hline\hline
Parameters & Solution I & Solution II & Solution III & Solution IV \\
\hline
$\Gamma_{\EE}\BR[\psi(3770)\to\ppjpsi]$ & \multicolumn{4}{c}{$0.5\pm0.1$~($0.4\pm0.1$)} \\
$\Gamma_{\EE}\BR(R_1\to\ppjpsi)$ & $8.8^{+1.5}_{-2.2}$~($\cdots$) & $6.8^{+1.1}_{-1.5}$~($\cdots$) & $7.2^{+0.9}_{-1.5}$~($\cdots$) & $5.6^{+0.6}_{-1.0}$~($\cdots$) \\
$\Gamma_{\EE}\BR(R_2\to\ppjpsi)$ & $13.3\pm1.4$~($12.0\pm1.0$) & $9.2\pm0.7$~($8.9\pm0.6$) & $2.3\pm0.6$~($2.1\pm0.4$) & $1.6\pm0.4$~($1.5\pm0.3$) \\
$\Gamma_{\EE}\BR(R_3\to\ppjpsi)$ & $21.1\pm3.9$~($17.9\pm3.3$) & $1.7^{+0.8}_{-0.6}$~($1.1^{+0.5}_{-0.4}$) & $13.3^{+2.3}_{-1.8}$~($12.4^{+1.9}_{-1.7}$) & $1.1^{+0.4}_{-0.3}$~($0.8\pm0.3$) \\
$\phi_1$ & $-58\pm 11$~($-33\pm8$) & $-116^{+9}_{-10}$~($-81^{+7}_{-8}$) & $65^{+24}_{-20}$~($81^{+16}_{-14}$) & $8\pm 13$~($33\pm9$) \\
$\phi_2$ & $-156\pm5$~($-132\pm3$) & $68\pm24$~($107\pm20$) & $-115^{+11}_{-9}$~($-95^{+6}_{-5}$) & $110\pm16$~($144\pm14$) \\
\hline\hline
\end{tabular}
\end{table*}

As an alternative description of the data, we use an exponential~\cite{exp} 
to model the cross
section near 4~GeV as in Ref.~\cite{babarnew}, instead of the
resonance $R_1$. 
The fit results are shown as dashed lines in Fig.~\ref{xsec-fit}. 
This model also describes data very well. A $\chi^2$-test to the fit
quality gives $\chi^2/ndf=93.2/111$. Thus, the existence of a
resonance near 4~GeV, such as the resonance $R_1$ or the
$Y(4008)$ resonance~\cite{belley}, is not necessary to explain the
data. The fit has four solutions with equally good fit quality~\cite{multi-solution} and
identical masses and widths of the resonances (listed in
Table~\ref{fit-mass}), while the phases and the product of the electronic widths with
the branching fractions 
are different (listed in Table~\ref{para-fit}). We
observe the resonance $R_2$ 
and the resonance $R_3$ again. 
The statistical
significance of resonance $R_3$ in this model is estimated to be
$7.6\sigma$ (including systematic uncertainties)
[$\Delta(-2\ln\mathcal{L})=70.7$, $\Delta ndf=4$] using the same
method as above.

The systematic uncertainty for the cross section measurement
mainly comes from uncertainties in the luminosity, efficiencies, radiative correction,
background shape and branching fraction of $\jpsi\to\LL$. The integrated luminosities
of all the data sets are measured using large angle Bhabha scattering events,
with an uncertainty of 1\%~\cite{lum}. The uncertainty in the
tracking efficiency for high momentum leptons is 1\% per track.
Pions have momenta that range from 0.1 to 1.06~GeV/$c$, and their
momentum weighted tracking efficiency uncertainty is also 1\%
per track. For the kinematic fit, we use a similar method as in
Ref.~\cite{kinematic-fit} to improve the agreement of the $\chi^2$ distribution
between data and MC simulation, and the systematic uncertainty for
the kinematic fit is estimated to be 0.6\% (1.1\%) for $\MM$
($\EE$) events. For the MC simulation of signal events, we use
both the $\pi^\pm\z^\mp$ model~\cite{bellenew,zc3900,zc-cleo} and the phase
space model to describe the $\EE\to\ppjpsi$ process. The efficiency
difference between these two models is 3.1\%, which is taken as
systematic uncertainty due to the decay model.

The efficiency for the other selection criteria, the trigger
simulation, the event start time determination and the FSR
simulation are quite high ($>99\%$), and their systematic errors are
estimated to be less than 1\%. In the ISR correction procedure, we
iterate the cross section measurement until $(1+\delta)\epsilon$
converges. The convergence criterion is
taken as the systematic uncertainty due to the ISR correction, which is 1\%.
We obtain the number of signal events by either fitting or
counting events in the $M(\LL)$ distribution. The background shape
is described by a linear distribution. Varying the background
shape from a linear shape to a second-order polynomial causes a
1.6\% (2.1\%) difference for the $\jpsi$ signal yield for the $\MM$
($\EE$) mode, which is taken as the systematic uncertainty for
background shape. The branching fraction of $\jpsi\to \LL$ is
taken from PDG~\cite{pdg}, the errors are 0.6\% for both $\jpsi$
decay modes. Assuming all the sources of systematic 
uncertainty to be independent, the total systematic uncertainties
are obtained by adding them in quadrature, resulting in 5.7\% for the
$\MM$ mode, and 5.9\% for the $\EE$ mode.

In both fit scenarios to the $\EE\to \ppjpsi$ cross section, we
observe the resonance $R_2$ and $R_3$. 
Since we can not distinguish the two scenarios from data,
we take the difference in mass and width as the systematic
uncertainties, i.e. 1.1~(6.8)~MeV/$c^2$ for the mass and
0.0~(3.2)~MeV for the width of $R_2~(R_3)$. The absolute c.m.
energy of all the data sets were measured with dimuon events, with
an uncertainty of $\pm 0.8$~MeV. Such kind of common uncertainty
will propagate only to the masses of the resonances with the same
amount, i.e. $\pm 0.8$~MeV/$c^2$. In both fits, the $\psi(3770)$
amplitude was added incoherently. The possible interference effect
of $\psi(3770)$ component was investigated by adding it coherently in
the fit with various phase. The largest deviation of the
resonant parameters between the fits with and without interference
for the $\psi(3770)$ amplitude are taken as systematic error,
which is 0.3~(1.3)~MeV/$c^2$ for the mass, and 2.0~(9.7)~MeV for
the width of the $R_2~(R_3)$ resonance. Assuming all the
systematic uncertainties are independent, we get the total
systematic uncertainties by adding them in quadrature, which is
1.4~(7.0)~MeV/$c^2$ for the mass, and 2.0~(10.2)~MeV for the width
of $R_2~(R_3)$, respectively.

In summary, we perform a precise cross section measurement of
$\EE\to \ppjpsi$ for c.m.\ energies from $\sqrt{s}=3.77$ to
4.60~GeV. Two resonant structures are observed, one
with a mass of $(4222.0\pm 3.1\pm 1.4)$~MeV/$c^2$ and a width of
$(44.1\pm 4.3\pm 2.0)$~MeV, and the other with a mass of
$(4320.0\pm 10.4 \pm 7.0)$~MeV/$c^2$ and a width of
$(101.4^{+25.3}_{-19.7}\pm 10.2)$~MeV, where the first errors are
statistical and the second ones are systematic. The first resonance
agrees with the $Y(4260)$ resonance reported by
\emph{BABAR}, CLEO and Belle~\cite{babary,cleoy,belley,babarnew,bellenew}. 
However, our measured width is much narrower than the $Y(4260)$ 
average width~\cite{pdg} reported by previous experiments. This is thanks
to the much more precise data from BESIII, which results in the observation
of the second resonance. The second resonance 
is observed for the first time in the process $\EE\to\ppjpsi$.
Its statistical significance is estimated to be larger than
$7.6\sigma$. The second resonance has a mass and width
comparable to the $Y(4360)$ resonance reported by Belle and
\emph{BABAR} in $\EE\to \pp\psi(2S)$~\cite{y4360}. 
If we assume it is the same resonance as the
$Y(4360)$, we observe a new decay channel of $Y(4360)\to
\ppjpsi$ for the first time.
Finally, we can not confirm the existence of the $Y(4008)$ 
resonance~\cite{belley,bellenew} from our data, since a continuum term 
also describes the cross section near 4~GeV equally well.

The BESIII collaboration thanks the staff of BEPCII and the IHEP
computing center for their strong support. This work is supported in
part by National Key Basic Research Program of China under Contract
No. 2015CB856700; National Natural Science Foundation of China (NSFC)
under Contracts Nos. 11235011, 11322544, 11335008, 11425524; the
Chinese Academy of Sciences (CAS) Large-Scale Scientific Facility
Program; the CAS Center for Excellence in Particle Physics (CCEPP);
the Collaborative Innovation Center for Particles and Interactions
(CICPI); Joint Large-Scale Scientific Facility Funds of the NSFC and
CAS under Contracts Nos. U1232201, U1332201; CAS under Contracts
Nos. KJCX2-YW-N29, KJCX2-YW-N45; 100 Talents Program of CAS; National
1000 Talents Program of China; INPAC and Shanghai Key Laboratory for
Particle Physics and Cosmology; German Research Foundation DFG under
Contracts Nos. Collaborative Research Center CRC 1044, FOR 2359;
Istituto Nazionale di Fisica Nucleare, Italy; Joint Large-Scale
Scientific Facility Funds of the NSFC and CAS	under Contract
No. U1532257; Joint Large-Scale Scientific Facility Funds of the NSFC
and CAS under Contract No. U1532258; Koninklijke Nederlandse Akademie
van Wetenschappen (KNAW) under Contract No. 530-4CDP03; Ministry of
Development of Turkey under Contract No. DPT2006K-120470; NSFC under
Contract No. 11275266; The Swedish Resarch Council; U. S. Department
of Energy under Contracts Nos. DE-FG02-05ER41374, DE-SC-0010504,
DE-SC0012069, DESC0010118; U.S. National Science Foundation;
University of Groningen (RuG) and the Helmholtzzentrum fuer
Schwerionenforschung GmbH (GSI), Darmstadt; WCU Program of National
Research Foundation of Korea under Contract No. R32-2008-000-10155-0.


\section*{Appendix}

\begin{table*}
\begin{center}
\caption{The c.m. energy ($\sqrt{s}$), integrated luminosity ($\mathcal{L}$),
number of $\jpsi$ signal events ($N^{\rm sig}$), detection efficiency ($\epsilon$),
radiative correction factor ($1+\delta$) and measured cross section [$\sigma(\EE\to\ppjpsi)$]
of ``XYZ data". The first errors are statistical and the second ones are systematic.} \label{xsec-xyz}
\begin{tabular}{cccccc}
\hline\hline
$\sqrt{s}$~(GeV) & $\mathcal{L}$ (pb$^{-1}$) & $N^{\rm sig}$ & $\epsilon$ & $1+\delta$ & $\sigma({\rm pb})$ \\
  \hline
  3.7730 & 2931.8 &       $3093.3\pm61.5$                    &      0.423      & 0.732 & $28.5\pm0.6\pm1.7$ \\
  3.8077 & 50.5     &          $34.7\pm6.9$                       &      0.396      & 0.871 & $16.7\pm3.3\pm1.0$ \\
  3.8962 & 52.6     &          $36.1\pm7.1$                       &      0.393      & 0.856 & $17.1\pm3.4\pm1.0$ \\
  4.0076 & 482.0   &         $325.8\pm21.7$                    &      0.392      & 0.901 & $16.0\pm1.1\pm1.0$ \\
  4.0855 & 52.6     &          $33.9\pm6.9$                       &      0.374      & 0.961 & $15.0\pm3.1\pm0.9$ \\
  4.1886 & 43.1     &          $26.9\pm6.5$                       &      0.394      & 0.858 & $15.5\pm3.8\pm0.9$ \\
  4.2077 & 54.6     &          $114.9\pm11.6$                   &      0.446      & 0.740 & $53.4\pm5.4\pm3.1$ \\
  4.2171 & 54.1     &          $130.5\pm12.2$                   &      0.458      & 0.731 & $60.3\pm5.7\pm3.5$ \\
  4.2263 & 1091.7 &         $3853.1\pm68.1$                  &      0.465      & 0.748 & $85.1\pm1.5\pm4.9$ \\
  4.2417 & 55.6     &          $203.5\pm15.1$                   &      0.453      & 0.802 & $84.4\pm6.3\pm4.9$ \\
  4.2580 & 825.7   &        $2220.9\pm53.7$                   &      0.444      & 0.853 & $59.5\pm1.4\pm3.4$ \\
  4.3079 & 44.9     &         $101.7\pm11.2$                    &      0.398      & 0.917 & $52.0\pm5.7\pm3.0$ \\
  4.3583 & 539.8   &       $621.5\pm28.8$                      &     0.372       & 1.022 & $25.4\pm1.2\pm1.5$ \\
  4.3874 & 55.2     &         $50.5\pm8.1$                        &     0.331       & 1.155 & $20.0\pm3.2\pm1.2$ \\
  4.4156 & 1073.6 &        $574.5\pm28.3$                     &     0.302       & 1.227 & $12.1\pm0.6\pm0.7$ \\
  4.4671 & 109.9   &         $63.4\pm9.8$                        &     0.293       & 1.240 & $13.3\pm2.1\pm0.8$ \\
  4.5271 & 110.0   &         $50.0\pm8.8$                        &     0.293       & 1.223 & $10.6\pm1.9\pm0.6$ \\
  4.5745 & 47.7     &         $26.1\pm6.1$                        &     0.281       & 1.213 & $13.4\pm3.2\pm0.8$ \\
  4.5995 & 566.9   &       $143.4\pm15.9$                      &     0.274       & 1.205 & $6.4\pm0.7\pm0.4$ \\
  \hline\hline
\end{tabular}
\end{center}
\end{table*}

\begin{table*}
\begin{center}
\caption{The c.m. energy ($\sqrt{s}$) and measured cross section [$\sigma(\EE\to\ppjpsi)$]
of ``Scan data". The first errors are statistical and the second ones are systematic.} \label{xsec-scan}
\begin{tabular}{ccccccccc}
\hline\hline
$\sqrt{s}$~(GeV) & $\sigma$~(pb) & $\sqrt{s}$~(GeV) & $\sigma$~(pb) & $\sqrt{s}$~(GeV) & $\sigma$~(pb) & $\sqrt{s}$~(GeV) & $\sigma$~(pb) \\
\hline
$3.8874$ & $9.7^{+13.1}_{-9.1}\pm0.6$ & $3.8924$ & $14.3^{+13.8}_{-9.8}\pm0.8$ & $3.8974$ & $20.7^{+13.4}_{-9.3}\pm1.2$ & $3.9024$ & $18.5^{+14.2}_{-10.2}\pm1.1$\\
$3.9074$ & $16.0^{+12.8}_{-8.5}\pm0.9$ & $3.9124$ & $12.2^{+13.4}_{-9.2}\pm0.7$ & $3.9174$ & $3.6^{+11.4}_{-6.6}\pm0.2$ & $3.9224$ & $26.9^{+17.1}_{-12.6}\pm1.6$\\
$3.9274$ & $24.2^{+15.6}_{-11.1}\pm1.4$ & $3.9324$ & $6.8^{+12.4}_{-8.1}\pm0.4$ & $3.9374$ & $13.5^{+12.7}_{-8.5}\pm0.8$ & $3.9424$ & $17.1^{+12.6}_{-8.6}\pm1.0$\\
$3.9474$ & $22.2^{+14.8}_{-11.0}\pm1.3$ & $3.9524$ & $18.0^{+13.0}_{-9.3}\pm1.0$ & $3.9574$ & $21.0^{+13.9}_{-10.4}\pm1.2$ & $3.9624$ & $15.5^{+12.3}_{-8.4}\pm0.9$\\
$3.9674$ & $14.4^{+13.2}_{-9.1}\pm0.8$ & $3.9724$ & $9.9^{+12.7}_{-9.0}\pm0.6$ & $3.9774$ & $9.2^{+11.7}_{-7.9}\pm0.5$ & $3.9824$ & $25.2^{+14.5}_{-10.8}\pm1.5$\\
$3.9874$ & $10.0^{+12.1}_{-8.4}\pm0.6$ & $3.9924$ & $1.0^{+10.5}_{-6.6}\pm0.1$ & $3.9974$ & $18.5^{+12.7}_{-8.9}\pm1.1$ & $4.0024$ & $21.2^{+14.9}_{-11.0}\pm1.2$\\
$4.0074$ & $21.0^{+14.3}_{-10.2}\pm1.2$ & $4.0094$ & $10.4^{+13.3}_{-8.9}\pm0.6$ & $4.0114$ & $25.0^{+15.3}_{-10.9}\pm1.4$ & $4.0134$ & $13.3^{+13.8}_{-9.2}\pm0.8$\\
$4.0154$ & $14.8^{+13.6}_{-9.3}\pm0.9$ & $4.0174$ & $36.5^{+17.2}_{-13.0}\pm2.1$ & $4.0224$ & $32.7^{+16.6}_{-12.2}\pm1.9$ & $4.0274$ & $9.1^{+7.7}_{-6.0}\pm0.5$\\
$4.0324$ & $22.3^{+15.2}_{-10.9}\pm1.3$ & $4.0374$ & $-2.4^{+11.7}_{-7.0}\pm0.1$ & $4.0474$ & $-1.2^{+12.6}_{-8.0}\pm0.1$ & $4.0524$ & $24.8^{+14.4}_{-10.2}\pm1.4$\\
$4.0574$ & $14.7^{+13.9}_{-9.2}\pm0.9$ & $4.0624$ & $13.3^{+13.4}_{-9.2}\pm0.8$ & $4.0674$ & $10.7^{+12.3}_{-8.2}\pm0.6$ & $4.0774$ & $19.1^{+13.6}_{-9.9}\pm1.1$\\
$4.0874$ & $12.2^{+12.9}_{-9.1}\pm0.7$ & $4.0974$ & $7.5^{+11.7}_{-7.6}\pm0.4$ & $4.1074$ & $9.9^{+12.6}_{-8.5}\pm0.6$ & $4.1174$ & $7.2^{+11.2}_{-7.3}\pm0.4$\\
$4.1274$ & $10.0^{+12.7}_{-8.5}\pm0.6$ & $4.1374$ & $29.8^{+15.1}_{-11.1}\pm1.7$ & $4.1424$ & $12.4^{+12.5}_{-8.6}\pm0.7$ & $4.1474$ & $9.5^{+11.4}_{-7.3}\pm0.6$\\
$4.1574$ & $29.4^{+15.5}_{-11.8}\pm1.7$ & $4.1674$ & $6.8^{+6.5}_{-4.8}\pm0.4$ & $4.1774$ & $26.0^{+14.4}_{-10.2}\pm1.5$ & $4.1874$ & $4.4^{+11.2}_{-7.0}\pm0.3$\\
$4.1924$ & $27.7^{+14.6}_{-10.6}\pm1.6$ & $4.1974$ & $35.3^{+15.5}_{-11.5}\pm2.0$ & $4.2004$ & $49.1^{+19.9}_{-15.6}\pm2.8$ & $4.2034$ & $26.4^{+15.9}_{-11.9}\pm1.5$\\
$4.2074$ & $29.7^{+15.1}_{-11.1}\pm1.7$ & $4.2124$ & $69.2^{+19.8}_{-16.1}\pm4.0$ & $4.2174$ & $64.3^{+19.5}_{-15.9}\pm3.7$ & $4.2224$ & $83.7^{+20.0}_{-16.6}\pm4.9$\\
$4.2274$ & $124.5^{+22.9}_{-19.7}\pm7.2$ & $4.2324$ & $69.4^{+18.2}_{-15.0}\pm4.0$ & $4.2374$ & $99.4^{+21.4}_{-18.0}\pm5.8$ & $4.2404$ & $74.7^{+18.3}_{-15.2}\pm4.3$\\
$4.2424$ & $47.0^{+15.5}_{-12.3}\pm2.7$ & $4.2454$ & $60.5^{+16.5}_{-13.5}\pm3.5$ & $4.2474$ & $66.3^{+16.6}_{-13.5}\pm3.8$ & $4.2524$ & $45.7^{+14.7}_{-11.7}\pm2.7$\\
$4.2574$ & $75.9^{+17.1}_{-14.3}\pm4.4$ & $4.2624$ & $58.2^{+15.9}_{-12.9}\pm3.4$ & $4.2674$ & $75.6^{+17.2}_{-14.3}\pm4.4$ & $4.2724$ & $53.0^{+16.0}_{-13.0}\pm3.1$\\
$4.2774$ & $38.4^{+14.1}_{-11.0}\pm2.2$ & $4.2824$ & $60.5^{+16.6}_{-13.6}\pm3.5$ & $4.2874$ & $60.1^{+15.7}_{-12.8}\pm3.5$ & $4.2974$ & $32.4^{+14.3}_{-11.1}\pm1.9$\\
$4.3074$ & $64.0^{+16.4}_{-13.3}\pm3.7$ & $4.3174$ & $39.1^{+13.3}_{-10.4}\pm2.3$ & $4.3274$ & $27.9^{+13.2}_{-10.0}\pm1.6$ & $4.3374$ & $31.0^{+13.3}_{-10.2}\pm1.8$\\
$4.3474$ & $14.0^{+11.4}_{-8.2}\pm0.8$ & $4.3574$ & $37.5^{+14.8}_{-11.6}\pm2.2$ & $4.3674$ & $34.8^{+13.7}_{-10.6}\pm2.0$ & $4.3774$ & $17.1^{+12.2}_{-8.9}\pm1.0$\\
$4.3874$ & $20.5^{+13.2}_{-9.6}\pm1.2$ & $4.3924$ & $23.8^{+13.2}_{-9.5}\pm1.4$ & $4.3974$ & $17.5^{+12.1}_{-8.2}\pm1.0$ & $4.4074$ & $4.7^{+11.0}_{-6.2}\pm0.3$\\
$4.4174$ & $16.9^{+12.3}_{-8.6}\pm1.0$ & $4.4224$ & $19.1^{+12.4}_{-8.6}\pm1.1$ & $4.4274$ & $9.9^{+11.9}_{-7.6}\pm0.6$ & $4.4374$ & $18.7^{+12.1}_{-8.4}\pm1.1$\\
$4.4474$ & $3.0^{+10.2}_{-6.4}\pm0.2$ & $4.4574$ & $6.9^{+9.4}_{-6.1}\pm0.4$ & $4.4774$ & $12.2^{+11.2}_{-7.7}\pm0.7$ & $4.4974$ & $1.0^{+8.3}_{-4.3}\pm0.1$\\
$4.5174$ & $12.7^{+10.2}_{-6.7}\pm0.7$ & $4.5374$ & $13.6^{+10.6}_{-7.5}\pm0.8$ & $4.5474$ & $14.7^{+10.8}_{-7.4}\pm0.9$ & $4.5574$ & $4.9^{+10.0}_{-6.2}\pm0.3$\\
$4.5674$ & $7.8^{+10.6}_{-6.8}\pm0.5$ & $4.5774$ & $8.7^{+11.1}_{-7.5}\pm0.5$ & $4.5874$ & $2.0^{+8.7}_{-4.4}\pm0.1$\\

\hline\hline
\end{tabular}
\end{center}
\end{table*}


\begin{thebibliography}{**}
\bibitem{exp} $p_0e^{-p_1(\sqrt{s}-M_{\rm th})}\Phi(\sqrt{s})$,
where $p_0$ and $p_1$ are free parameters, $M_{\rm
th}=2m_\pi+m_{\jpsi}$ is the mass threshold of the $\ppjpsi$ system, and
$\Phi(\sqrt{s})$ the phase space factor.
\end{thebibliography}
\end{document}